\documentclass[aps, reprint, groupedaddress, showpacs, superscriptaddress, amsmath, amssymb]{revtex4-1} 
\usepackage[colorlinks,allcolors=blue]{hyperref}
\usepackage{graphicx}
\usepackage{dcolumn}
\usepackage{bm}
\usepackage{braket}

\begin{document}
    \title{Dirac Cone Pairs in Silicene Induced by Interface Si-Ag Hybridization: A First Principles Effective Band Study}
    \author{Chao Lian}
    \affiliation{Beijing National Laboratory for Condensed Matter Physics and Institute of Physics, Chinese Academy of Sciences, Beijing, 100190, P. R. China}
    
    \author{Sheng Meng}
    \email{smeng@iphy.ac.cn}
    \affiliation{Beijing National Laboratory for Condensed Matter Physics and Institute of Physics, Chinese Academy of Sciences, Beijing, 100190, P. R. China}
    \affiliation{Collaborative Innovation Center of Quantum Matter, Beijing, 100190, P. R. China}
    
    \begin{abstract}
        Using density functional theory combined with orbital-selective band unfolding techniques, we study the effective band structure of silicene ($3\times3$)/Ag(111) ($4\times4$) structure. Consistent with the ARPES spectra recently obtained by Feng {\it et al.} [Proc. Natl. Acad. Sci. 113, 14656 (2016)], we observe six pairs of Dirac cones 
        near the boundary of the Brillouin zone (BZ) of Ag(1$\times$1), while no Dirac cone is observed inside the BZ. 
        Furthermore, we find that these Dirac cones are induced by the interfacial Si-Ag hybridization, mainly composed of Si $p_z$ orbitals and Ag $sp$ bands, which is intrinsically different from the Dirac cones in free-standing silicene.
    \end{abstract}
    
    \maketitle
    \section{Introduction}
    Silicene is a promising candidate material for developing next generation field effect transistors~\cite{Tao2015, Jia2015,drummond2012electrically,liu2013bilayer,Sakai2015,Calixto2015,MehdiAghaei2015,du2014tuning,Nigam2015,liu2014band, doi:10.1021/nl203065e, Li2012, Lian2013, Lian2015}, valleytronic devices~\cite{Wang2015h, Vargiamidis2015, Zhai2016, Li2017, Yang2015b, Missault2016, Qu2016, Zhang2016b, Qiu2015b, PingNiu2015, Luo2016, Wang2015i, Zhang2016c, Khani2016, ezawa2012valley, Qiu2016, Hajati2016, Missault2015, Azarova2017, BinHo2016, Wu2015a} and quantum spin Hall devices~\cite{PhysRevLett.107.076802, Geissler2013, Yao2015a, Cao2015, liu2011low, Zhang2016a}, thanks to its relatively strong spin-orbital coupling and compatibility with silicon based technology. These applications are closely associated with the presence of Dirac electrons in free-standing silicene predicted from theory~\cite{cahangirov2009two}. 
    
    Although free-standing silicene possess Dirac cone band structures, the existence of Dirac cones in supported silicene is heavily under debate in experiment, especially for the most common system experimentally probed: silicene synthesized on Ag(111)~\cite{Vogt2012, chen2012evidence, feng2012evidence, jamgotchian2012growth, lin2012structure, feng2013observation, majzik2013combined, resta2013atomic, johnson2014metallic, Mahatha2014, Wang2016d, Takagi2015, tchalala2014atomic, liu2014various, Du2016}. {Due to the strong Si-Ag interaction, different silicene structures are formed on the Ag substrate, such as 3$\times$3, $\sqrt{3}\times\sqrt{3}$, $2\sqrt{3}\times2\sqrt{3}$ and $\sqrt{17}\times\sqrt{17}$. Among these phases, the most common silicene structures are the 3$\times$3 phase and the $\sqrt{3}\times\sqrt{3}$ phase, which can be grown by changing the substrate temperature~\cite{feng2012evidence}. {Although the $3\times3$ phase forms on a Ag(111) substrate, as more Si atoms are deposited the structure reconstructs into the $\sqrt{3}\times\sqrt{3}$ phase and forms multilayers. Therefore, although $3\times3$ can be regarded as stable at low temperatures, it is an intermediate phase and it eventually reconstructs into the $\sqrt{3}\times\sqrt{3}$ phase in ambient conditions. The $\sqrt{3}\times\sqrt{3}$ phase has the lowest energy per surface area,  which also suggests that it is the most stable phase at high Si coverage. It also has better agreement with experimental observations of multilayer silicene.~\cite{Cahangirov2014, Cahangirov2014a}} The atomistic structures of these two phases are studied with density functional calculations~\cite{Vogt2012, chen2012evidence, Fu2015b, Cahangirov2014, Cahangirov2014a}. Since the 3$\times$3 phase (abbreviated as Si/Ag hereafter) is commonly synthesized and generally accepted, whether the Dirac cone exists in 3$\times$3 phase is quite important but still under debate.} Several experiments including scanning tunneling spectroscopy (STS)~\cite{chen2012evidence, feng2013observation} and angle-resolved photoemission spectroscopy (ARPES) studies~\cite{Vogt2012} suggest that the Dirac cone structure is preserved in Si/Ag. In contrast, density functional theory (DFT) and other experimental studies~\cite{PhysRevLett.110.076801, PhysRevB.87.245430, Chen2014a, Mahatha2014} claim that the opposite is true: the Dirac cone is absent in Si/Ag, due to the strong Si-Ag interaction and significant charge transfer between Si and Ag layers. 
    
    \begin{figure}
        \includegraphics[width=\linewidth]{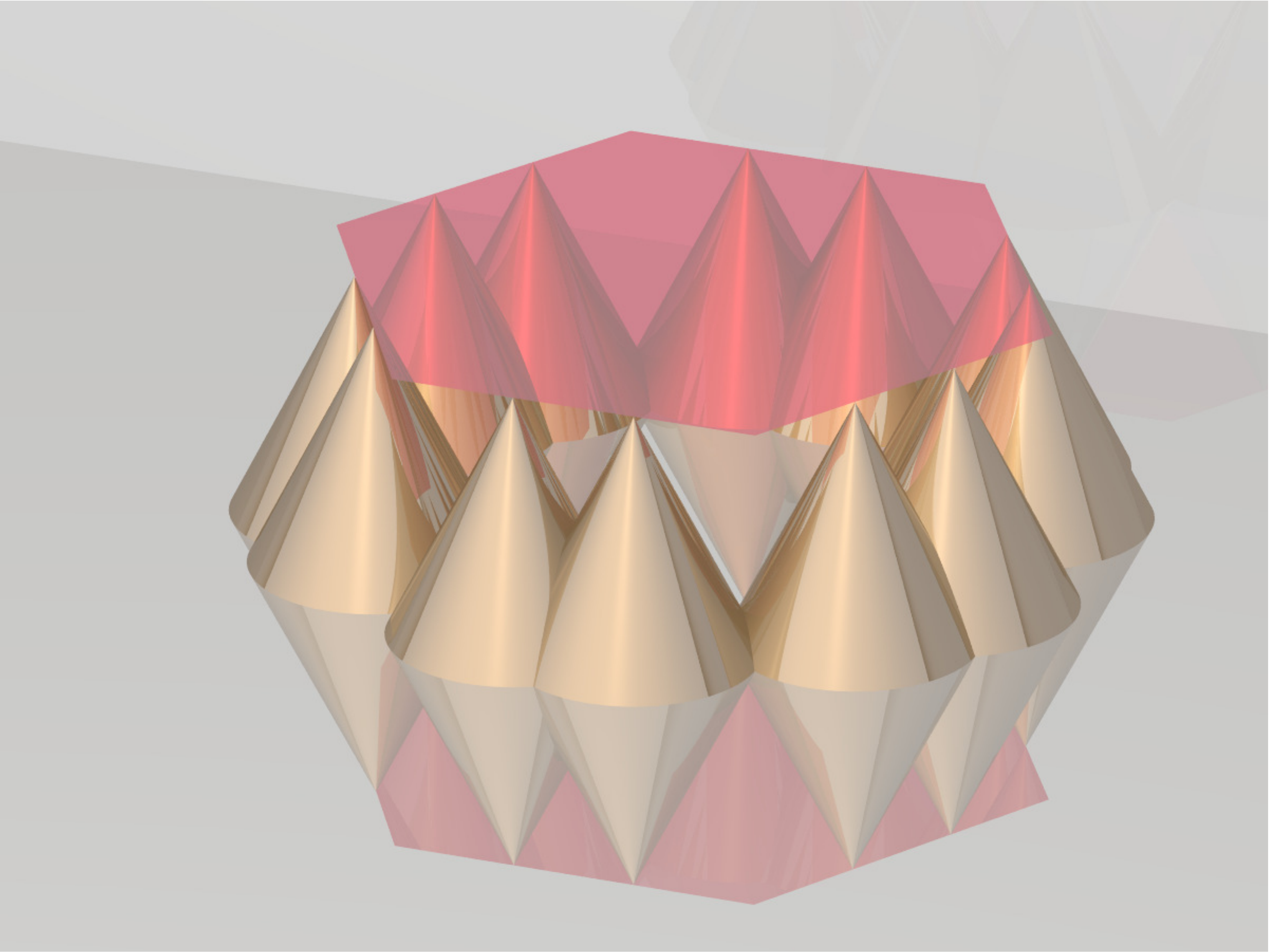}
        \caption{\label{fig:model} Schematic illustration of the cone pairs in silicene monolayer supported on Ag(111). The red plane represents the first Brillouin zone of Ag($1\times1$).}. 
    \end{figure}
    
    To maintain Dirac cone band structures, efforts are being made to peel the free-standing silicene off the metal substrate where silicene was originally grown. However, directly removing the Ag substrate is technically difficult~\cite{Tao2015}. Quite surprisingly, Feng {\it et al.} recently reported Dirac cone pairs in Si/Ag~\cite{Feng2015}. In their ARPES measurements, six pairs of Dirac cones (Fig.~\ref{fig:model}) are observed at the edge of the Ag(1$\times$1) Brillouin zone (BZ) [Fig.~\ref{struct_and_BZ}(c)]. The Dirac cone pairs remain even with the presence of the Ag substrate. This study not only proposes a recipe to achieve silicene-based high speed electronic devices, but also suggests mechanisms of the formation of Dirac cone electronic structures. Unfortunately, the authors pointed out that the experimental observation can-not be explained in terms of existing band structure calculations. In addition, the underlying mechanism of the formation of Dirac cone structures in Si/Ag is still illusive. 
    
    In the present work we reproduce the ARPES observation of Dirac cones based on first-principles DFT calculations combined with orbital-selective band unfolding techniques. We demonstrate the presence of six pairs of Dirac cones near the boundary of the Brillouin zone (BZ) of Ag(1$\times$1). 
    Our theoretical results are highly consistent with experimental data presented in the ARPES study, except that we find evidence for the upper branches of Dirac cones being absent in both intact and doped systems. We find that the Dirac cones are not the intrinsic properties of either silicene itself or the underlying Ag slab. Instead, these Dirac cones are induced by the strong interface Si-Ag hybridization, mainly composed of Si $p_z$ orbitals and Ag $sp$ bands, which are radically different from the Dirac cones of free-standing silicene.
    
    \section{Methods}
    To reveal the key factors underlying the ARPES experiments, first-principles DFT calculations are performed to reproduce the ARPES spectra. We note that the energy bands from regular DFT calculations are different from the ARPES spectra in three aspects: \\
    (i) For the systems containing translational symmetry breaking, e.g. reconstructions or impurities, a supercell (SC) approach is usually adopted in DFT band calculations. Then DFT bands are folded into the supercell Brillouin zone, namely the Brillouin zone associated with the supercell symmetry,  while the ARPES measurements still span over the primitive cell (PC) Brillouin zone. \\
    (ii) The DFT bands are often referred to the energy dispersion relation $E(\mathbf{k})$. However the ARPES spectra are related to the spectral function $A(\mathbf{k},\varepsilon)$, which is reduced to $E(\mathbf{k})$ only in the picture of the single-particle Green function.\\
    (iii) The DFT bands always comprise all electronic states of the cell under computation, including contributions from both the surface and the underlying substrate. By contrast, ARPES is mainly a surface sensitive technique, only the electronic states near the surface contribute to the ARPES spectra. 
    
    To bridge the gap between DFT bands and measured ARPES spectra, band unfolding is used to calculate the effective band structure (EBS) of the SC~\cite{Popescu2012,Medeiros2014}, in response to statement (i) and (ii). With respect to (iii), an extra weight function $W(N,\mathbf{K})$ is invoked to realize the orbital selection rules. By selecting the orbitals of surface atoms, we obtain the EBS originated from the selected surface atoms, which are directly comparable to the measured ARPES spectra.
    
    We rewrite the electronic wavefunction $\ket{\Psi_{N,\mathbf{K}}}$ from SC calculations in the basis of PC wavefuncitons $\ket{\psi_{n,\mathbf{k}_i}}$. Here we use capital letters to denote quantities associated with SC and lowercase letters for quantities associated with PC:
    \begin{equation}
    \ket{\Psi_{N,\mathbf{K}}}=\sum_{n,\mathbf{k}_i}a(n,\mathbf{k}_i;N,\mathbf{K})\ket{\psi_{n,\mathbf{k}_i}},
    \end{equation}
    where $N$ denotes the band index and $\mathbf{k}_i=\mathbf{K}+\mathbf{G}$ ($\mathbf{G}$ is the reciprocal vector of SC). Using the Bl\"och theorem and the plane-wave basis, $\ket{\psi_{n,\mathbf{k}_i}}$ and $\ket{\Psi_{N,\mathbf{K}}}$ can be written as:
    \begin{equation}
    \label{Bloch_PC}
    \begin{split}
    \ket{\psi_{n,\mathbf{k}_i}} &= {u_{n,\mathbf{k}_i}(\mathbf{r})} \exp(i\mathbf{k}_i\cdot\mathbf{r}) \\
    & =\left[\sum_{\mathbf{g}}{c_{n,\mathbf{k}_i}(\mathbf{g})} \exp(i\mathbf{g}\cdot\mathbf{r})\right]\exp(i\mathbf{k}_i\cdot\mathbf{r}),
    \end{split}
    \end{equation}
    \begin{equation}
    \label{Bloch_SC}
    \begin{split}
    \ket{\Psi_{N,\mathbf{K}}} 
    &={U_{N,\mathbf{K}}(\mathbf{R})}\exp(i\mathbf{K}\cdot\mathbf{R}) \\
    &=\left[\sum_{\mathbf{G}}{C_{N,\mathbf{K}}(\mathbf{G})}\exp(i\mathbf{G}\cdot\mathbf{R})\right]\exp(i\mathbf{K}\cdot\mathbf{R}),
    \end{split}
    \end{equation}
    where $\ket{u(\mathbf{r})}$ is the Bl\"och function and $C(\mathbf{G})$ is the coefficient of the Bloch function on the plane-wave basis.
    The spectral function is:
    \begin{equation}
    A(\mathbf{k}_i,\varepsilon)=\sum_N P(\mathbf{k}_i;\mathbf{K},N)\delta(\varepsilon-\varepsilon(N,\mathbf{K})),
    \end{equation}
    where
    \begin{equation}
    \label{probability}
    \begin{split}
    P(\mathbf{k}_i;\mathbf{K},N)&=\sum_{n}a^*(\mathbf{k}_i,n;\mathbf{K},N)a(\mathbf{k}_i,n;\mathbf{K},N) \\
    &=\sum_{n}\braket{\Psi_{N,\mathbf{K}} | \psi_{n,\mathbf{k}_i}}\braket{\psi_{n,\mathbf{k}_i}| \Psi_{N,\mathbf{K}}} \\
    &=\sum_{\mathbf{g}}\left|C_{N,\mathbf{K}}(\mathbf{g}+\mathbf{k}_i-\mathbf{K})\right|^2.
    \end{split}
    \end{equation}
    In the derivation of Eq.~\ref{probability}, Eqs.~(\ref{Bloch_PC}) and (\ref{Bloch_SC}) are used. Details can be found in Ref.~\cite{Popescu2012}. 
    Since only $C_{N,\mathbf{K}}$ is needed in Eq.~(\ref{probability}), only the SC wavefunction is calculated.
    
    To achieve a selection of orbitals from specific atoms, we introduce an extra weight function $W(N,\mathbf{K})$ in calculating the spectral function, which is modified as  
    \begin{equation}
    \label{spacial_band_unfolding}
    A(\mathbf{k},\varepsilon)=\sum_N P(\mathbf{k}_i;\mathbf{K},N) W(N,\mathbf{K}) \delta(\varepsilon-\varepsilon(N,\mathbf{K})).
    \end{equation} 
    In principle, the choice of $W(N,\mathbf{K})$ is arbitrary. Here, we set $W(N,\mathbf{K})$ to be the projected density of states (PDOS) of certain orbitals $D(N,\mathbf{K},\{\eta_i\})$. Thus, the spectral function in Eq.~(\ref{spacial_band_unfolding}) describes the EBS contributed by the orbitals in $\{\eta_i\}$. Our code is based on the BandUP code~\cite{Medeiros2014,Medeiros2015} and we modify it to include the partial projections. 
    
    The first principles calculations are performed with the Vienna Ab initio Simulation Package (VASP)~\cite{PhysRevB.47.558, PhysRevB.49.14251, PhysRevB.54.11169}. The projector augmented-waves method~\cite{PhysRevB.50.17953} and Perdew-Burke-Ernzerhof exchange correlation~\cite{Perdew1996} are used. The plane-wave cutoff energy is set to be 250~eV. The vacuum space is set to be larger than 15 {\AA}. The Brillouin zone is sampled using a Monkhorst-Pack scheme~\cite{monkhorst1976special}. We use a \textit{k}-point mesh of $6\times6\times1$ for structural optimization and $12\times12\times1$ in the self-consistent calculations. Using the conjugate gradient method, the positions of atoms are optimized until the convergence of the force on each atoms is less than 0.005~eV/{\AA}.

    \section{Results and Discussion}
    \begin{figure}
        \includegraphics[width=\linewidth]{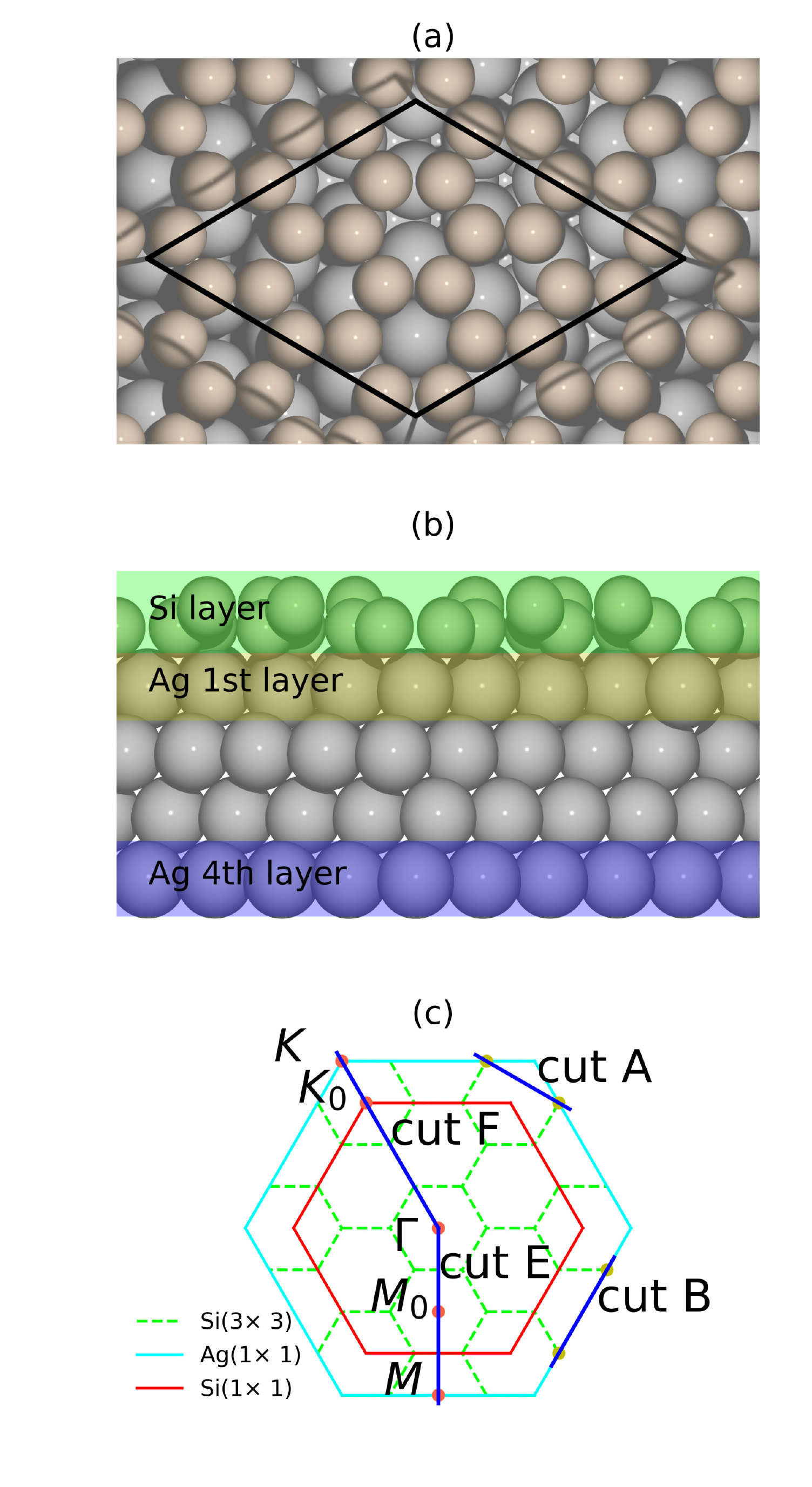}
        \caption{\label{struct_and_BZ} (a) Top view and (b) side view of the structure of Si/Ag. The brown balls denote the silicon atoms and the grey balls denote the silver atoms. (c) The diagram of the Brillouin zones of the Si($3\times3$), Si(1$\times$1) and Ag($1\times1$). {Yellow dots denote the Dirac cones along cut A and cut B.}}
    \end{figure}
    
    \begin{figure*}
        \includegraphics[width=\linewidth]{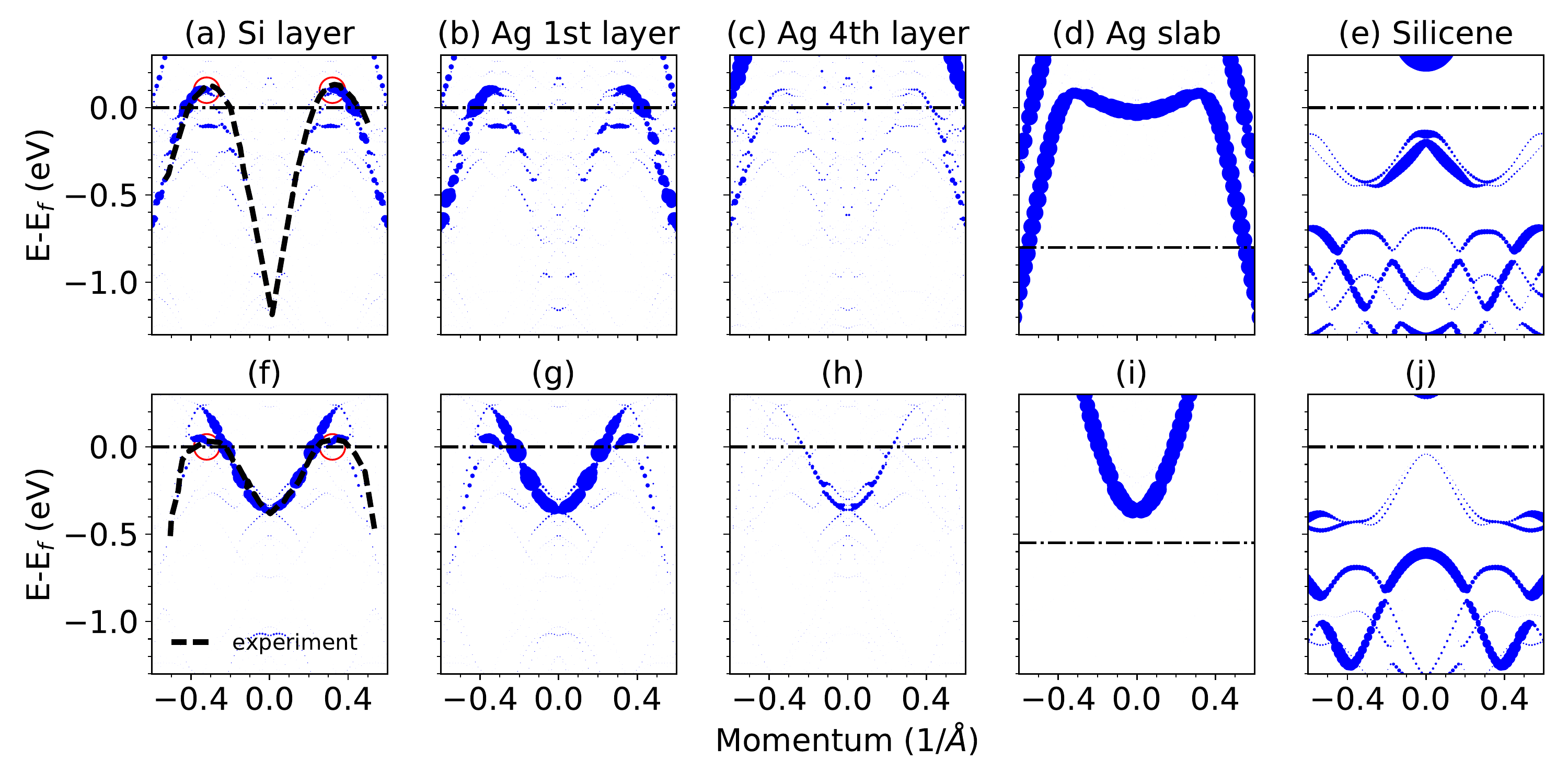}
        \caption{EBS along cut A (a--e) and cut B (f--j) of Si/Ag projected on different layers, Ag slab and silicene. The ARPES spectra~\cite{Feng2015}, shown with dashed lines, are upward shifted for direct comparison. {The red circles denote the positions of the Dirac cones.}}
        \label{fig:cone}
    \end{figure*}
    
    As shown in Fig.~\ref{struct_and_BZ}, {We model silicene supported on Ag(111) with a four-layer Ag slab covered by a monolayer of silicene. The initial structure is set to be 3$\times$3 reconstructed phase manually. After geometry relaxation, the Si layer on the Ag slab exhibits 3$\times$3 periodicity with respect to the free standing silicene $1\times1$ structure with atomic reconstructions. The lattice constants are $a = b = 11.76$~\AA, $c = 27.57$~\AA\, and $\alpha=60$, $\beta= 90$, $\gamma = 90$. The average Si-Ag bond distance is about 2.7~\AA, lying in between the values for the Ag-Ag bond length of 2.92~\AA and the Si-Si bond length of 2.36~\AA.} It indicates that a strong Si-Ag interaction could take place once Si atoms are deposited onto Ag(111). Thus, three major ingredients can alter the electronic properties in Si/Ag as compared to freestanding silicene: (i) the $3\times3$ reconstruction of silicene; (ii) the electron transfer between Si layer and Ag substrate; and (iii) the orbital hybridization between Si and Ag atoms. This complexity leads to the possibility of forming Dirac cones of different origins: (i) Dirac cones coming from the bare reconstructed silicene with possible doping; (ii) from bare Ag substrate; (iii) from band renormalization induced by strong Si-Ag hybridization. 
    
    
    To explore the underlying mechanism of the Dirac cone pairs observed in ARPES, we first isolate the contributions of Si and Ag to the band structure of the composite system, by projecting the EBS on different layers of Si/Ag and comparing to those for isolated silicene and the Ag(111) slab. The EBS of Si/Ag are projected on the Si layer, and on the first layer and the fourth layer of the Ag(111) slab, as shown in Fig.~\ref{struct_and_BZ}(c). Here we denote the EBS of $X$ ($X =$ Si/Ag, silicene, or Ag slab) projected onto the $Y$ atomic layer ($Y =$ Si, Ag first layer, or Ag fourth layer) as EBS($Y$@$X$). The difference between EBS(Si layer@Si/Ag) and EBS(silicene) is that the former includes the influence from the Ag substrate while there is none in the latter. Similarly, the comparison between EBS(Ag layer@Si/Ag) and EBS(Ag slab) reveals the effect of the Si layer on the Ag slab.
    
    \begin{figure}
        \centering
        \includegraphics[width=1.0\linewidth]{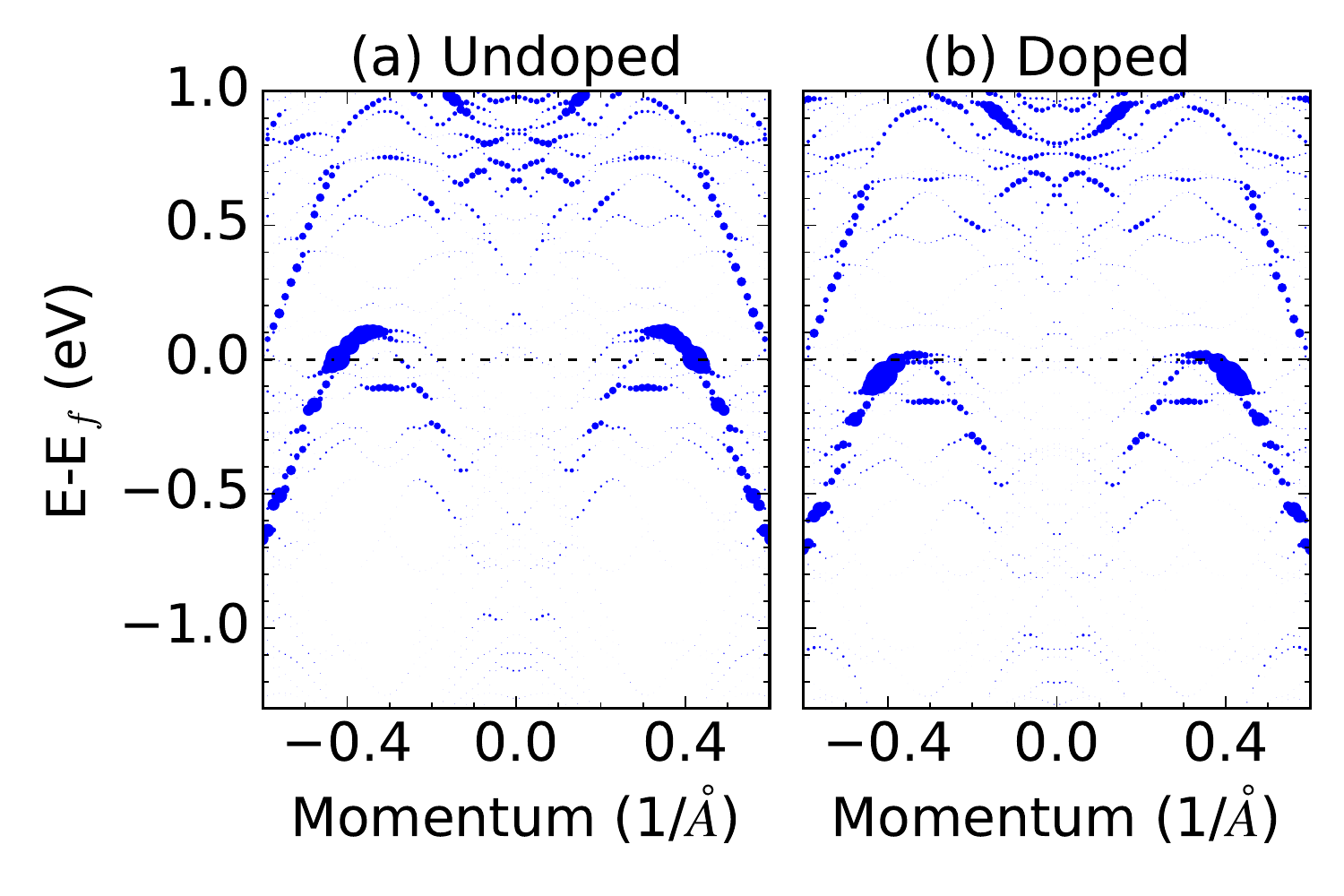}
        \caption{EBS(Si layer@Si/Ag) of (a) undoped Si/Ag, (b) Si/Ag doped with one potassium atom per unit cell.}
        \label{fig:doping}
    \end{figure}

    \begin{figure*}
        \includegraphics[width=\linewidth]{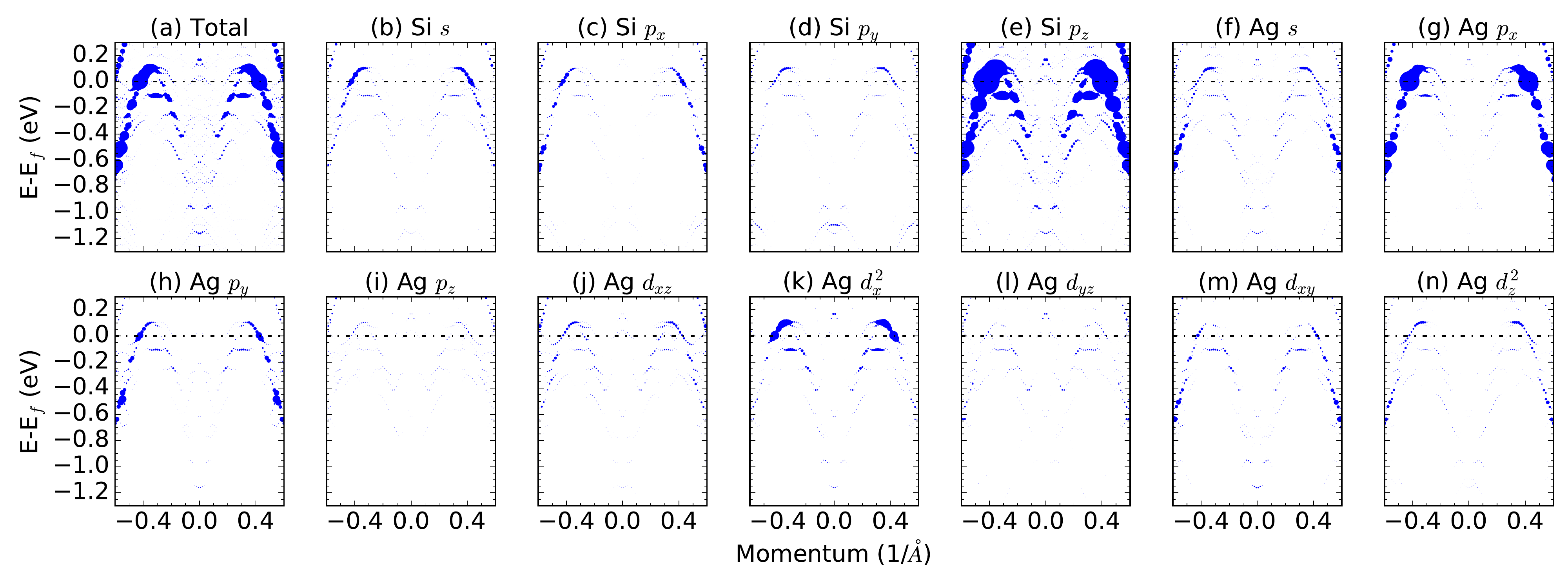}
        \caption{\label{fig:orbital} EBS(Si/Ag) along cut A projected on different orbitals.}
    \end{figure*}
    
    \begin{figure*}
        \includegraphics[width=\linewidth]{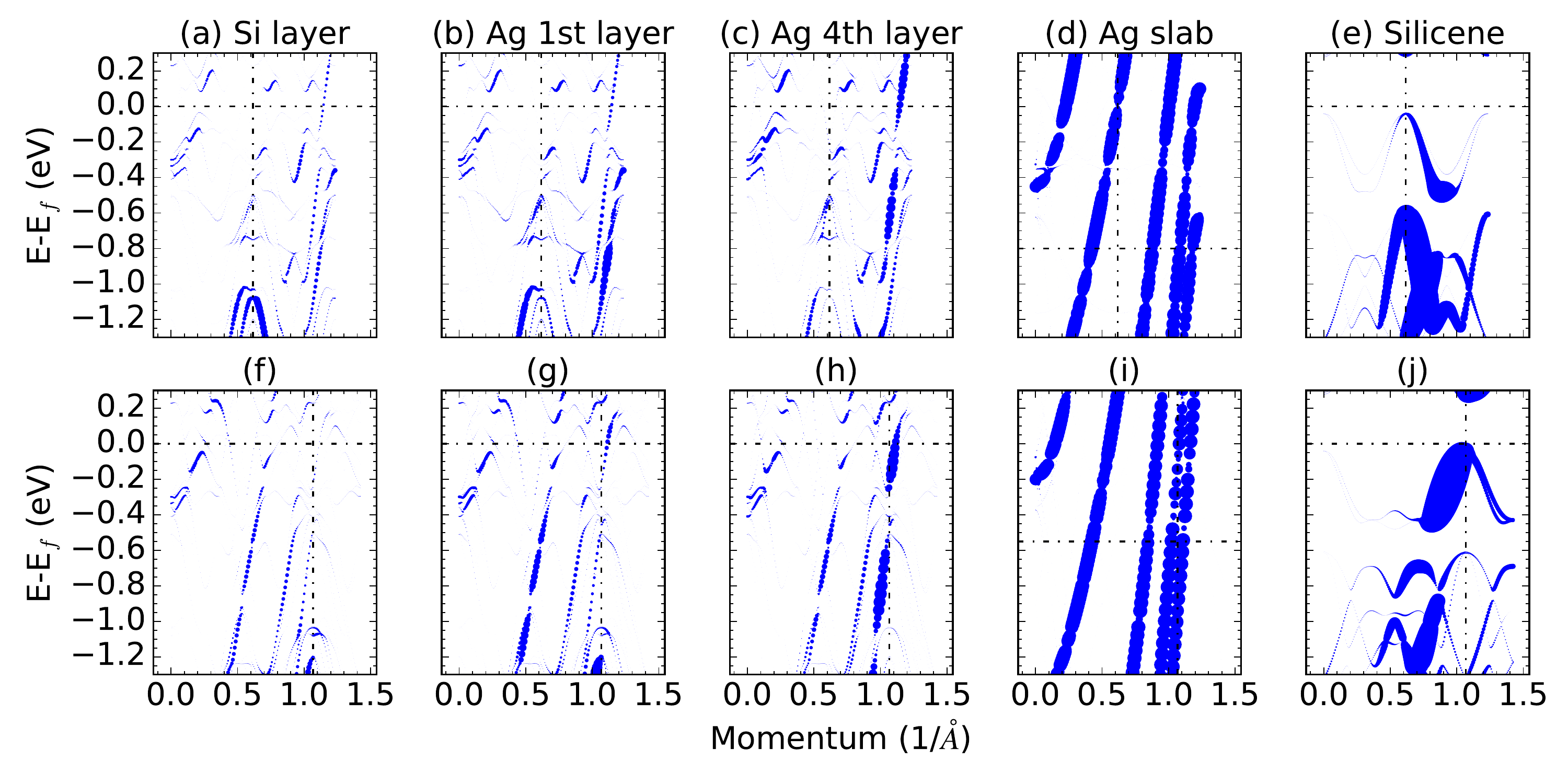}
        \caption{EBS along cut E (a--e) and cut F (f--j) of Si/Ag projected on different layers, Ag slab and silicene. Verticle lines show the position of $M_0$ and $K_0$ in cuts E and F, respectively.}
        \label{fig:cross}
    \end{figure*}
    The cone pairs reported in ARPES locate at the edge of the BZ of Ag($1\times1$), which can be measured directly along cut A and B in Fig.~\ref{struct_and_BZ}(c), the same cut as those used in the experiment. The EBS along cut A and B are shown in Fig.~\ref{fig:cone}.
    
    Figure~\ref{fig:cone}(a) and (f) shows the EBS(Si layer@Si/Ag) along cut A and cut B, respectively. The  spectra simulated for the EBS(Si layer@Si/Ag) using band unfolding techniques described above show an excellent agreement with the experimental ARPES spectra. Both theoretical and experimental spectra show an evident peak-valley-peak feature in Fig.~\ref{fig:cone}(a). The peaks are claimed to come from a pair of Dirac cones in the ARPES measurement~\cite{Feng2015}. The features of EBS(Si layer@Si/Ag), EBS(Ag slab) and EBS(silicene) are summarized in Table~\ref{tab:EBS}.
    
    \begin{table} 
        \caption{\label{tab:EBS} The features of EBS(Si layer@Si/Ag), EBS(Ag slab) and EBS(silicene).}
        \begin{ruledtabular}
            \begin{tabular}{lccc}
                EBS & Si layer@Si/Ag(\AA) &Ag slab &Silicene \\\hline
                Cut A & peak-valley-peak & plateau & peak \\
                Cut B & peak-valley-peak & valley  & peak \\ 
                Cut E & no feature & no feature & cones\\
                Cut F & no feature & no feature & cones\\
            \end{tabular}
        \end{ruledtabular}
    \end{table}

    The Dirac cones in silicene/Ag(111) are radically different from the Dirac cones in free-standing silicene in two major aspects. First, these cones in Si/Ag are not strictly linear in energy dispersion, which is akin to a gapped cone. Nevertheless, the effective electron masses near the cones are calculated to be very small, only $4.3\times 10^{-3} m_e$ fitted to the data in Fig.~\ref{fig:cone}(a), where $m_e$ is the mass of a free electron. With the presence of such low-mass quasiparticles, it is a promising material to build high-speed electronic devices. Second, the upper branches of the cones are absent. As shown in Figs.~\ref{fig:cone}(a)-\ref{fig:cone}(f), the upper branch of the Dirac cones brought by strong Si-Ag hybridzation is not clearly visible in our theoretical band analysis. 
    It might be shifted to higher energy above the Fermi level or is further mixed with other Si/Ag bands, thus being hidden in the background of unoccupied effective bands. {Motivated by experiments~\cite{Feng2015}, the upper branch of the Dirac cone pairs may be tuned by potassium doping. Thus, we dope one potassium atom in the Si(3$\times$3)/Ag($4\times4$) supercell. The K-K distance is 11.76~\AA, sufficiently large to avoid the K-K interaction. The EBS is only downshifted by 0.15~eV without modifications to its overall shape (Fig.~\ref{fig:doping}), indicating pure electron doping induced by potassium.}
    Thus, strictly speaking, these Dirac cones are Dirac-like gapped half cones. Note that these Dirac cone pairs only exist in $3\times3$ phase. {No Dirac cone pair is observed in our EBS study of the of $\sqrt{3}\times\sqrt{3}$ phase~\cite{Fu2015b}. }
    
    To reveal their origin, we find that the Dirac cone features in the EBS plot of Si/Ag come from strong Si-Ag hybridization. Since the weight function in Eq.~\ref{spacial_band_unfolding} is the PDOS of the selected orbitals, the similarity between EBS(Si layer@Si/Ag) and EBS(Ag first layer@Si/Ag) indicates a strong hybridization between the Si layer and the first Ag layer. The Dirac cones become less obvious in the EBS(Ag fourth layer@Si/Ag), which is similar to EBS(Ag slab), due to the weaker interaction between the Si layer and the fourth Ag layer. Our detailed analysis also indicate that the Dirac cones are primarily contributed by Si $p_z$ orbitals and its hybridization with Ag $sp$ orbitals~[Fig.~\ref{fig:orbital}]. 
    
    In addition, we note that the EBS(Ag slab) is largely upwards shifted for comparison, which indicates a strong electron transfer from Si layer to Ag slab in the supported monolayer silicene on Ag(111). Therefore, the observed Dirac cones in Si/Ag are the combined result of hybridization and electron transfer between the Si layer and the Ag layer.
    
    These Dirac cones do not originate solely from the bare silicene or bare Ag slab. First, since the intrinsic Dirac cone is folded onto the $\Gamma$ point in the BZ of Si($3\times3$), it is counter-intuitive that only six pairs of Dirac cones are observed at the edge of the Ag($1\times1$) BZ, instead of at the edge or center of the BZ for Si($1\times1$) or Si($3\times3$). It is predictable that there are gapped cones in EBS(silicene) at the the middle of cut B, the M and M$_0$ along cut E, and at K$_0$ along cut F. Our calculations of EBS(silicene) verify this prediction [Figs.~\ref{fig:cone}(j), \ref{fig:cross}(e), and \ref{fig:cross}(j)]. However, strongly influenced by the Ag slab, no Dirac cone is observed at these points in both experimental and our simulated ARPES spectra. 
    
    We note EBS(Si layer@Si/Ag) are radically different from EBS(silicene). At $k=0$ of cut B, EBS(Si layer@Si/Ag) is valley-like while EBS(silicene) presents a peak-like feature. This Dirac cones at M, M$_0$ and K$_0$ point are also absent in EBS(Si layer@Si/Ag). Thus, the cones in EBS(Si layer@Si/Ag) are different from the gapped cone in EBS(silicene).
    
    Second, there is no cone in EBS(Ag slab) along cuts A and B. 
    {As shown in Figs.~\ref{fig:cone}(d) and ~\ref{fig:cone}(j) and Table~\ref{tab:EBS}, the EBS(Ag slab) shows a plateau along cut A and a single valley along cut B.} Thus, the cones in EBS(Si layer@Si/Ag) are absent in EBS(Ag slab). Moreover, the Ag slab has a strong signal near the Fermi level along cuts E and F. EBS signals are weak in EBS(Si layer@Si/Ag), and become stronger when the Ag contribution increases (Ag fourth layer@Si/Ag layer). These bands are similar to the EBS of bare silver slab. The similarity between the EBS of Si/Ag and that of a bare Ag slab shows that the EBS of Si/Ag are dominated by the Ag substrate, but also are being strongly modified by the presence of the Si layer. The screening of the Si layer explains the obscure signals along cut E in experimental ARPES results.

    \section{Conclusion}
    Using the density functional theory combined with the orbital-selective band unfolding technique, we study the effective band structures (EBS) of Si($3\times3$)/Ag($4\times4$). Consistent with the ARPES measurement recently reported by Feng {\it et al.}~\cite{Feng2015}, we observe six pairs of Dirac cones near the boundary of the Brillouin zone (BZ) of Ag(1$\times$1), while no Dirac cone is observed inside the BZ. We find that Dirac cones are not the intrinsic properties of the silicene or the Ag slab; instead, these Dirac cones are emergent phenomena induced by the strong Si-Ag hybridization; they are composed of Si $p_z$ orbitals and Ag $sp$ orbitals, radically different from the Dirac cones of free-standing silicene. This study clarifies the nature of Dirac electrons in the composite silicene/Ag(111) system, and hints that a range of new quasiparticles and emergent phenomena could be employed by delicate interface engineering. 
    
    \section{Acknowledgments}
    We acknowledge financial supports from MOST (Grants No. 2016YFA0300902 and No. 2015CB921001), NSFC (Grant No. 11474328), and CAS (XDB07030100). We acknowledge helpful discussions with Prof. Xing-Jiang Zhou.

\end{document}